\documentclass{article}

\usepackage[english]{babel}
\usepackage{comment}

\usepackage[letterpaper,top=2cm,bottom=2cm,left=3cm,right=3cm,marginparwidth=1.75cm]{geometry}

\usepackage{amsmath}
\usepackage{graphicx}
\usepackage[colorlinks=true, allcolors=blue]{hyperref}

\title{Practical and Ready-to-Use Methodology to Assess the re-identification Risk in Anonymized Datasets}
\author{Louis-Philippe SONDECK, Maryline LAURENT}

\begin{document}
\date{} 
\maketitle

\begin{abstract}
To prove that a dataset is sufficiently anonymized, many privacy policies suggest that a re-identification risk assessment be performed, but do not provide a precise methodology for doing so, leaving the industry alone with the problem.
This paper proposes a practical and ready-to-use methodology for re-identification risk assessment, the originality of which is manifold: (1) it is the first to follow well-known risk analysis methods (e.g. EBIOS) that have been used in the cybersecurity field for years, which consider not only the ability to perform an attack, but also the impact such an attack can have on an individual; (2) it is the first to qualify attributes and values of attributes with e.g. degree of exposure, as known real-world attacks mainly target certain types of attributes and not others.


\textit{\textbf{Keywords} Privacy Impact Assessment, re-identification risk assessment, Anonymized dataset, Dataset, GDPR, HIPAA, Privacy.}

\end{abstract}

\section{Introduction}
There are several policies in the healthcare industry that address the issue of anonymizing data for disclosure, including the Health Insurance Portability and Accountability Act (HIPAA) Privacy Rule \cite{HIPAA}, the EMA 0070 Policy \cite{EMA}, and the GDPR \cite{GDPR2018}. These guidelines generally refer to well-known anonymization techniques (e.g. k-anonymity, l-diversity, differential privacy,...). Many of them define a two-sided approach to anonymization, either by applying a well-defined method (e.g. Safe Harbor for HIPAA, WP29 anonymization criteria fulfillment for GDPR \cite{WP29Opinion}) or by performing a re-identification risk assessment (e.g. Expert Determination for HIPAA, risk analysis assessment for GDPR). However, the well-defined methods are in most cases either too restrictive, leading to a poor useful dataset (e.g. GDPR, EMA Policy 0070) or too simplistic, leading to privacy concerns (e.g. HIPAA). 

This paper proposes a practical and ready-to-use methodology for re-identification risk assessment and results in quantifying the risk to privacy. It emphasizes the notion of "feared event", which is well known in the context of cybersecurity risk, but which is not considered in current methodological proposals. This allows for a more accurate assessment of the re-identification risk, by taking into account not only the likelihood of an attack, which refers to the ability of an attacker to perform an attack, but also the severity of this attack, which describes the impact on the individual. The re-identification risk is then calculated based on both the likelihood and the severity. After positioning our contribution against current approaches in Section \ref{sec:definition}, three sections describe the method to compute severity (Section \ref{sec:severity}), likelihood (Section \ref{sec:likelihood}) and re-identification risk (Section \ref{sec:computing_rr}). Well-chosen examples are given throughout the paper to help understand the proposed methodology: the initial dataset in Table \ref{tab:initial_table} and the two anonymized versions of it in Tables \ref{tab:3-anon} and \ref{tab:hipaa}, which re-identification risk is assessed.

\begin{table}
    \centering
    \begin{tabular}{|c|c|c|c|c|c|c|}
        \hline
        \# & Age & Gender & Country & Admission Date & Blood Type & Disease\\\hline
        1 & 23 & M & Nigeria & 2019-09-21 & A+ & Colds\\
        2 & 23 & M & Cameroon & 2019-06-05 & O+ & Colds\\
        3 & 25 & M & Nigeria & 2019-06-05 & O+ & Colds\\
        4 & 32 & F & France & 2022-10-12 & O+ & Colds\\
        5 & 31 & F & France & 2022-10-12 & O+ & Flu\\
        6 & 37 & F & Spain & 2022-05-14 & AB+ & HIV\\
        7 & 51 & F & Canada & 2021-04-01 & AB- & Diabetes\\
        8 & 53 & F & USA & 2021-04-01 & O- & Cancer\\
        9 & 53 & F & Mexico & 2021-04-01 & B- & HIV\\
        10 & 57 & F & Canada & 2022-04-01 & B- & Colds\\
        11 & 36 & F & Belgium & 2017-02-12 & O+ & Flu\\
        12 & 30 & F & Italy & 2015-01-02 & AB- & Flu\\\hline
    \end{tabular}
    \caption{Initial Dataset}
    \label{tab:initial_table}
\end{table}

\begin{table}
    \centering
    \begin{tabular}{|c|c|c|c|c|c|c|}
        \hline
        \# & Age & Gender & Country & Admission Date & Blood Type & Disease\\\hline
         & 2* & M & Africa & 2019-**-** & Positive Rehsus & Colds\\
        Group 1 & 2* & M & Africa & 2019-**-** & Positive Rehsus & Colds\\
         & 2* & M & Africa & 2019-**-** & Positive Rehsus & Colds\\\hline
         & 3* & F & Europe & 2022-**-** & Positive Rehsus & Colds\\
        Group 2 & 3* & F & Europe & 2022-**-** & Positive Rehsus & Flu\\
         & 3* & F & Europe & 2022-**-** & Positive Rehsus & HIV\\\hline
         & 5* & F & America & 2021-**-** & Negative Rehsus & Diabetes\\
        Group 3 & 5* & F & America & 2021-**-** & Negative Rehsus & Cancer\\
         & 5* & F & America & 2021-**-** & Negative Rehsus & HIV\\\hline
    \end{tabular}
    \caption{Anonymized Dataset with k-anonymity (k=3)}
    \label{tab:3-anon}
\end{table}

\begin{table}
    \centering
    \begin{tabular}{|c|c|c|c|c|c|c|}
        \hline
        \# & Age & Gender & Country & Admission Date & Blood Type & Disease\\\hline
        1 & 23 & M & Nigeria & 2019-**-** & A+ & Colds\\
        2 & 23 & M & Cameroon & 2019-**-** & O+ & Colds\\
        3 & 25 & M & Nigeria & 2019-**-** & O+ & Colds\\
        4 & 32 & F & France & 2022-**-** & O+ & Colds\\
        5 & 31 & F & France & 2022-**-** & O+ & Flu\\
        6 & 37 & F & Spain & 2022-**-** & AB+ & HIV\\
        7 & 51 & F & Canada & 2021-**-** & AB- & Diabetes\\
        8 & 53 & F & USA & 2021-**-** & O- & Cancer\\
        9 & 53 & F & Mexico & 2021-**-** & B- & HIV\\
        10 & 57 & F & Canada & 2022-**-** & B- & Colds\\
        11 & 36 & F & Belgium & 2017-**-** & O+ & Flu\\
        12 & 30 & F & Italy & 2015-**-** & AB- & Flu\\\hline
    \end{tabular}
    \caption{HIPAA Anonymized Dataset}
    \label{tab:hipaa}
\end{table}

\section{Background and Positioning against EU and US Legislations}
\label{sec:definition}



\subsection{Re-identification Risk Calculated based on two Criteria: Severity and Likelihood}
\label{sec:compute-introduction}

In cybersecurity, every risk assessment begins by identifying the "feared event" that describes what is actually feared and should be avoided. In the context of personal data, the "feared event" is of 3 categories: illegitimate access, unwanted modification, and disappearance of data \cite{georgiou2023dpia} \cite{seyyar2020privacy}. For re-identification, the "feared event" is simply "illegitimate access", as we are trying to prevent sensitive data from being disclosed to unauthorized persons. Once identified, the "feared event" is then assessed to provide a "severity level", which is then used to calculate the risk. For re-identification, the "feared event" is assessed by the "Severity" of the impact on individuals, which depends on the sensitivity of the data to be disclosed (Section \ref{sec:severity}). To explain the relevance of the "Severity" of an attack for re-identification risk assessment, let us consider the dataset in Table \ref{tab:initial_table}. Although it is relatively easy to re-identify User 3, because he is unique within the table (the only one who is 25), the related feared event would be the disclosure of information "Colds", which is not very sensitive compared to "HIV"; therefore, the risk related to User 3 should be lower, compared to User 6, for example. However, "Severity" is only one part of the risk calculation, the other part is "Likelihood", which refers to the attackers' ability to execute an attack. Unlike the common definition of re-identification risk, which considers only the attackers' ability \cite{benitez2010evaluating} \cite{carey2023measuring} \cite{dankar2012estimating} \cite{el2011systematic}, cybersecurity risk considers both "Severity" and "Likelihood".

Drawing on proven cybersecurity risk analyses, e.g. EBIOS \cite{EBIOS}, we define the 2 vulnerabilities severity and likelihood, and the re-identification risk as follows: 
\begin{itemize}
    \item \textbf{Severity (S)}: Severity is related to the "feared event" and describes the impact on individuals if the "feared event" occurs. Severity in the context of re-identification reflects the sensitivity of the attributes to be disclosed and therefore, the impact on the individual if these sensitive attributes are disclosed.
    \item \textbf{Likelihood (L)}: Likelihood is the probability that an attack will occur. The likelihood describes the attacker's ability to perform an attack and how vulnerabilities can be exploited to do so. It is assessed by the \textbf{exploitability} of such vulnerabilities or by previous occurrences of such attacks. In the context of re-identification, the vulnerabilities can come from information inside the anonymized dataset (e.g. quasi-identifiers) and/or outside the anonymized dataset (e.g. flaws related to access controls). The exploitability describes how vulnerabilities can be used to link an individual to sensitive information.
    \item \textbf{Re-identification Risk (R)}: Based on the EBIOS definition of risk, we define re-identification risk as the probability of associating sensitive information with an individual and having an impact on that individual. It is calculated based on the severity of such an impact on the individual, and the likelihood of the attack. R can be calculated as follows: \textbf{R = S x L}.
\end{itemize}


Another point of interest in cybersecurity risk analyses, and one from which we can draw inspiration, is that it is not necessary to address all risks, but only the most important ones. Thanks to our methodology of assessing severity and likelihood at the attribute value level, this allows us to focus our attention on specific parts of the data, as shown in bold in Table \ref{tab:risky_entries}.

\subsection{Positioning the Contribution against Existing EU and US Legislations}
\label{sec:positioning}

The EU and US approaches to re-identification have some limitations; some are specific to each and others are common. Next, we position our contribution against both approaches.

\subsubsection{US Approach Limitations}
\label{sec:us_limitations}

The risks considered by the US approach \cite{el2008protecting} \cite{dankar2010method} are:

\begin{itemize}
    \item \textbf{Prosecutor risk}: the probability that a subject is unique in the dataset for an attacker who knows that the subject is in the dataset. 
    \item \textbf{Journalist risk}: the probability that a subject is unique in the population from which the dataset was sampled. The attacker does not know if the subject is in the dataset and assumes that if he is unique in the population from which the dataset was sampled, then he is unique in the dataset.
    \item \textbf{Marketer risk}: the probability that each subject is unique in the dataset, the attacker is willing to identify as many subjects as possible. The risk of the journalist or the prosecutor is always greater than or equal to the marketer risk.
\end{itemize}
\newpage
There are two main problems with the US approach to re-identification attacks. They do not allow a realistic assessment:
\begin{itemize}
    \item \textbf{The blur around the background knowledge}: to compute the risk of the journalist's attack, we need to consider a population (a background knowledge dataset), since the attacker does not know whether his target is in the anonymized dataset, and assumes that if the target is unique in the population, then the target is also unique in the anonymized dataset. 
  However, the problem is how to correctly guess which population the attacker is using, given that any dataset from any source can be used, as long as it contains some of the attributes corresponding to those in the anonymized dataset, and the attacker has reason to believe that his target is within that dataset. It is therefore very difficult to quantify such a risk related to very relative information. 
  
  \textbf{=$>$ Our proposed methodology introduces a new criterion "Exposure" to quantify the risk related to a background knowledge}, which is based on the type of attributes and their ability to be found in other datasets (Section \ref{sec:exposure}).

    \item \textbf{Re-identification occurs only with uniqueness}: the US approach considers that re-identification can only occur with uniqueness, which is not the case. In fact, the prosecutor's attack and the journalist's attack allow to compute the re-identification risk based on the equivalence class within the anonymized dataset (prosecutor attack) or within the population dataset (journalist attack). Indeed, the re-identification score is computed as the inverse of the size of the equivalence class, which is actually the probability related to unique values within the equivalence class \cite{el2008protecting}. The marketer risk uses a similar approach. However, there are cases where uniqueness is not necessary for re-identification. To illustrate, consider Table \ref{tab:3-anon} which is a 3-anonymity table and by definition, prevents uniqueness. However, due to the lack of diversity for Group 1, it is enough to know that the target individual is in his twenties, to deduce that he suffers from "Colds". Therefore, in this case, the attacker does not need to find a unique individual, he only needs to find someone in his twenties for a complete re-identification (linkage with the information "Colds"). 
    
    \textbf{=$>$ Our proposed methodology suggests a new definition of re-identification based on the ability to associate a target with an information of interest}, as the goal of the attack is not only to make the target unique among others, but also to associate it with an information of interest (cf. Section \ref{sec:compute-introduction}). 
    
\end{itemize}

\subsubsection{EU Approach Limitations}
\label{sec:eu_limitations}

The vulnerabilities considered by EU in the WP29 Opinion on Anonymisation Techniques \cite{WP29Opinion} are: 

\begin{itemize}
    \item \textbf{Singling out}: the ability to isolate some or all of the records that identify an individual in the dataset. For example, in Table \ref{tab:initial_table}, the value "37" of "Age" attribute isolates User 6 because he is the only one of that age in the dataset.
    \item \textbf{Linkability}: the ability to link at least two records concerning the same data subject or a group of data subjects (either in the same database or in two
different databases). For example, using the demographic attributes in Table \ref{tab:initial_table} ("Age", "Gender", "Country"), we can link this dataset to an auxiliary dataset containing the same attributes concerning the same individuals.
    \item \textbf{Inference}: the ability to deduce, with significant probability, the value of an attribute from the values of a set of other attributes. For example, in Table \ref{tab:initial_table}, all men "M" have "Colds".
\end{itemize}

We consider "Singling out" to be a special case of "Inference". In fact, "Singling out" is used to isolate an individual within the dataset, and this is a case of perfect inference of the sensitive information. For example, in Table \ref{tab:initial_table}, User 6 can be singled out by his age ("37"), and this is used to infer that he has "HIV". The more general case of inference is described in Table \ref{tab:3-anon} where inference is performed through equivalence classes. For example, using Group 1, we can perform a different case of perfect inference, as all the individuals in this equivalence class suffer from "Colds". The other equivalence classes do not allow a perfect inference as they are well diversified.

\textbf{=$>$ Therefore, our framework only considers "Inference" and "Linkability" for computing the likelihood of an attack }(see Section \ref{sec:likelihood}).

\subsubsection{Common Limitations of EU and US Approaches}
\label{sec:common_limitations}
Both approaches suffer from a mismatch between likelihood and re-identification.
In fact, as explained in Section \ref{sec:compute-introduction}, a risk is not only about the ability of an attacker to perform an attack (likelihood), but also about the sensitivity of the information to be disclosed (severity). Our framework takes both severity and likelihood to compute the re-identification risk (Section \ref{sec:severity} and Section \ref{sec:likelihood})




\section{Computing the Severity (S)}
\label{sec:severity}
Severity refers to the impact that disclosure of sensitive information could have on the individual, and this depends on the sensitivity of the attribute. For example, in Table \ref{tab:initial_table}, not all attributes have the same sensitivity because some may have more impact on the individual than others. For example, linking someone to a disease has a greater impact than linking them to their country or age as attribute "Disease" may reveal more compromising information about the individual. In addition, the difference in the sensitivity levels is also related to the specific values of the attributes. For example, for the "Disease" attribute, the value "Colds" is less sensitive than the values "HIV" or "Diabetes" because the latter two are serious diseases while the former is not. Our framework takes into account the severity of a risk based on the sensitivity of specific attributes and values.

To assess the sensitivity of a given attribute, we use the classification proposed by the French Data Protection Authority (CNIL) \cite{CNIL-PIA}, which defines 3 types of effects on an individual, with 4 levels for each: negligible (level 1), limited (level 2), significant (level 3), maximum (level 4). The 3 types of effects are: 
\begin{itemize}
\item Bodily: refers to the impacts related to the body of the individual (e.g. impairment of physical integrity as a result of an assault, domestic accident, work accident, etc.)
\item Material: refers to the effects related to the material assets of the individual (e.g. insurance price increase, bank ban)
\item Moral: refers to impacts related to moral issues (e.g. sense of invasion of privacy and irreparable harm, discrimination)
\end{itemize}

Based on this classification, we evaluate each of the attributes of Table \ref{tab:initial_table} with the results provided in Tables \ref{tab:attributes_severity_evaluation} and \ref{tab:values_severity_evaluation}. In Table \ref{tab:attributes_severity_evaluation}, only the attribute type is considered and a Global Severity level, ranging from 1 to 4, is calculated as the maximum of all sensitivity levels associated with each impact type. In Table \ref{tab:values_severity_evaluation}, we focus on the values of the "Disease" attribute and assign the most sensitivity to the values "HIV" and "Cancer", as the disclosure of this information may have a greater impact on the individual.
Based on these two tables, we can already propose a first assessment of the initial Table \ref{tab:initial_table}. The resulting Table \ref{tab:risky_entries} highlights the entries that can have the greatest impact on individuals.   

\begin{table}
    \centering
    \begin{tabular}{|c|c|c|c|c|}
        \hline
         Attributes & Bodily Impact & Material Impact & Moral Impact & Global Severity\\\hline
         Age  & 1-Negligible & 1-Negligible & 1-Negligible & 1-Negligible\\
         Gender & 1-Negligible & 1-Negligible & 1-Negligible & 1-Negligible \\
         Country  & 1-Negligible  & 1-Negligible & 1-Negligible & 1-Negligible\\
         Admission Date  & 1-Negligible & 1-Negligible & 1-Negligible & 1-Negligible\\
         Bloody Type & 1-Negligible & 1-Negligible & 1-Negligible & 1-Negligible \\
         \textbf{Disease} & 1-Negligible & \textbf{3-Significant} & \textbf{4-Maximum} & \textbf{4-Maximum}\\\hline
    \end{tabular}
    \caption{Attribute Severity Rating in the Initial Dataset}
    \label{tab:attributes_severity_evaluation}
\end{table}

\begin{table}
    \centering
    \begin{tabular}{|c|c|}
    \hline
        Values & Severity\\\hline
        Colds & 1-Negligible\\
        Flu & 1-Negligible\\
        \textbf{Diabetes} & \textbf{3-Significant}\\
        \textbf{HIV} & \textbf{4-Maximum}\\
        \textbf{Cancer} & \textbf{4-Maximum}\\\hline
    \end{tabular}
    \caption{Severity Rating of the Values carried by the Disease Attribute in the Initial Dataset}
    \label{tab:values_severity_evaluation}
\end{table}

\begin{table}
    \centering
    \begin{tabular}{|c|c|c|c|c|c|c|}
        \hline
        \# & \textbf{Age} & \textbf{Gender} & \textbf{Country} & Admission Date & Blood Type & Disease\\\hline
        1 & 23 & M & Nigeria & 2019-09-21 & A+ & Colds\\
        2 & 23 & M & Cameroon & 2019-06-05 & O+ & Colds\\
        3 & 25 & M & Nigeria & 2019-06-05 & O+ & Colds\\
        4 & 32 & F & France & 2022-10-12 & O+ & Colds\\
        5 & 31 & F & France & 2022-10-12 & O+ & Flu\\
        6 & \textbf{37} & \textbf{F} & \textbf{Spain} & \textbf{2022-05-14} & \textbf{ AB+} & \textbf{HIV}\\
        7 & \textbf{51} & \textbf{F} & \textbf{Canada} & \textbf{2021-04-01} & \textbf{AB-} & \textbf{Diabetes}\\
        8 & \textbf{53} & \textbf{F} & \textbf{USA} & \textbf{2021-04-01 }& \textbf{O-} & \textbf{Cancer}\\
        9 & \textbf{53} & \textbf{F} & \textbf{Mexico} & \textbf{2021-04-01} & \textbf{B- }& \textbf{HIV}\\
        10 & 57 & F & Canada & 2022-04-01 & B- & Colds\\
        11 & 36 & F & Belgium & 2017-02-12 & O+ & Flu\\
        12 & 30 & F & Italy & 2015-01-02 & AB- & Flu\\\hline
    \end{tabular}
    \caption{Entries at Risk in the Initial Dataset according to the Severity of Values carried by the Disease Attribute}
    \label{tab:risky_entries}
\end{table}

\section{Computing the Likelihood (L)}
\label{sec:likelihood}

As presented in Section \ref{sec:compute-introduction}, the Likelihood refers to an attacker's ability to perform a re-identification attack, and thus his ability to link a given piece of sensitive information to an individual. To perform this attack, the attacker may rely on vulnerabilities inside and/or outside the anonymized records. As concluded in Section \ref{sec:eu_limitations}, he can carry out an inference attack and/or a linkability attack.   


To mitigate the "blur around the background knowledge", as presented in Section \ref{sec:us_limitations}, our framework focuses only on vulnerabilities within the anonymized dataset. 
These vulnerabilities arise from the quasi-identifiers that can be used to associate an individual with the sensitive information. We compute the likelihood as the level of exploitability of these quasi-identifiers to link the sensitive information to an individual. For example, in Table \ref{tab:hipaa}, this would consist of assessing the exploitability of the attributes "Age", "Gender", "Country", "Admission Date", and "Blood Type" to link an individual to the "Disease" attribute. This requires the two following components to be calculated separately:

\begin{itemize}
    \item Exposure of Attributes (Section \ref{sec:exposure}): it evaluates the Linkability vulnerability by assessing how easy it is for an attacker to find a given quasi-identifier in an auxiliary dataset, for cross-referencing with the anonymized dataset.
    
    \item Inference Vulnerability (Section \ref{sec:inference}): it focuses on the anonymized dataset and evaluates how easy it is for the attacker to infer the values of the sensitive attribute using the quasi-identifiers.
\end{itemize}


\subsection{Exposure of Attributes (Linkability)}
\label{sec:exposure}
The linkability vulnerability from the anonymized dataset is measured by the ability of the attributes to be linked to an auxiliary dataset. In fact, not all attributes have the same linking potential; some are easier to use than others. For example, in Table \ref{tab:initial_table}, the quasi-identifiers "Age", "Gender", and "Country" are easier to use for linkage than the quasi-identifiers "Admission Date" and "Blood Type". Indeed, the former can easily be found in an auxiliary dataset, as they are demographic data. 

To capture those characteristics, we introduce a new criterion called "Exposure". Exposure characterizes the ability of an attribute to be found in an auxiliary dataset. We define 4 levels of Exposure: 
\begin{itemize}
    \item "Internal Restricted" (IR): Attributes related to specific operations that are not used outside of the context of those operations and are therefore difficult to find in another dataset (e.g. some medical test results, technical data, etc.).
    \item "Internal Extended" (IE): Attributes used by several services within an organization, but not used outside the context of that organization (e.g. personnel number, department name, etc.).
    \item "External Restricted" (ER): Attributes that can be found outside the organization but require in-depth research on social networks (e.g. smoker/non smoker, etc.). 
    \item "External Extended" (EE): Attributes that are widely used in different contexts and easily accessible from social networks or search engines (e.g. demographics, last name, first name, age, location data, etc.).
\end{itemize}

However, while the exposure levels highlight whether the attributes are widespread enough to perform a linking attack, they are not sufficient to capture the linkability vulnerability. Indeed, we should also consider the granularity of the attribute, which refers to the number of different values of the attribute. For example, in Table \ref{tab:initial_table}, while the attributes "Age" and "Gender" are both "EE", they do not share the same Linkability level because the attribute "Age" is more granular (has more values) than the attribute "Gender", and is therefore easier to use for a linkage attack. The granularity of an attribute is captured by the "Singling out" vulnerability, which is already included in the inference vulnerability (Section \ref{sec:eu_limitations}), so it does not need to be explicitly included in the Likelihood computation.


\begin{table}
    \centering
    \begin{tabular}{|c|c|}
        \hline
         Attributes & Exposure \\\hline
         \textbf{Age}  & 4-EE \\
         \textbf{Gender} & 4-EE \\
         \textbf{Country}  & 4-EE \\
         Admission Date  & 2-EI \\
         Bloody Type & 1-RI \\\hline
    \end{tabular}
    \caption{Attributes Exposure Assessment}
    \label{tab:attributes_exposure_evaluation}
\end{table}

\subsection{Assessment of the Inference Vulnerability}
\label{sec:inference}

The inference vulnerability is computed between the quasi-identifiers and the sensitive attribute to infer the values of the sensitive attribute using the quasi-identifiers. For example, in Table \ref{tab:initial_table}, the sensitive attribute is attribute "Disease", and using the attribute "Country", it is possible to guess some of the values of the attribute "Disease". For example, all people from "Nigeria" suffer from "Colds". 
Finally, we can combine different attributes to calculate the inference vulnerability. For example, all males "M" who are "23" suffer from "Colds" in Table \ref{tab:initial_table}.

To calculate the inference vulnerability, we use the Discrimination Rate\footnote{Discrimination Rate is based on the entropy and it quantifies the extent to which quasi-identifiers values can discriminate the values of a sensitive attribute, with results between 0 and 1. $DR=1$ means it is sufficient to know the quasi-identifier value to guess the sensitive value (e.g. Group 1 in Table \ref{tab:3-anon}). The DR also applies to combinations of quasi-identifier values.} (DR) metric \cite{Sondeck2017} \cite{sondeck2017semantic}, although other metrics may be used:

\begin{figure}
    \centering
    \includegraphics[width=0.75\linewidth]{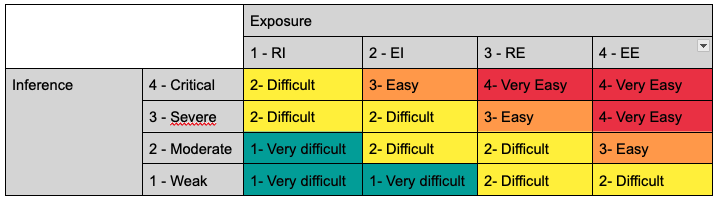}
    \caption{Our Exploitability Scale based on Exposure and Inference}
    \label{fig:exploitability_scale}
\end{figure}

\begin{enumerate}
    \item Weak: the DR is between 0 and 0.25
    \item Moderate: the DR is between 0.25 and 0.5
    \item Severe: the DR is between 0.5 and 0.75
    \item Critical: the DR is between 0.75 and 1
\end{enumerate}

\begin{table}
    \centering
    \begin{tabular}{|c|c|c|c|}
        \hline
         Combined Attributes & Exposure & Inference & Exploitability \\\hline
         \textbf{Age/Gender/Country}  & 4-EE & 4-Critical & \textbf{4-Very Easy} \\
         Admission Date/Blood Type & 2-EI & 4-Critical & 3-Easy\\\hline
    \end{tabular}
    \caption{Exploitability Assessment for the HIPAA Table w.r.t to  Quasi-identifiers}
    \label{tab:hipaa_attributes_exploitability_evaluation}
\end{table}

\begin{table}
    \centering
    \begin{tabular}{|c|c|c|c|}
        \hline
         Combined Attributes & Exposure & Inference & Exploitability \\\hline
         \textbf{Age/Gender/Country}  & 4-EE & 4-Critical & \textbf{4-Very Easy} \\
         Admission Date/Blood Type & 2-EI & 4-Critical & 3-Easy\\\hline
    \end{tabular}
    \caption{Exploitability Assessment for the k-anonymity Table w.r.t to Quasi-identifiers}
    \label{tab:3-anon_attributes_exploitability_evaluation}
\end{table}

\begin{table}
    \centering
    \begin{tabular}{|c|c|c|c|}
        \hline
         Attributes & Exposure & Inference & Exploitability \\\hline
         \textbf{Age}  & 4-EE & 4-Critical & \textbf{4-Very Easy} \\
         \textbf{Gender} & 4-EE & 3-Severe & \textbf{4-Very Easy}\\
         \textbf{Country}  & 4-EE & 4-Critical & \textbf{4-Very Easy}\\
         Admission Date  & 2-EI & 4-Critical & 3-Easy\\
         Blood Type & 2-EI & 3-Severe & 2-Difficult\\\hline
    \end{tabular}
    \caption{Exploitability Assessment for the k-anonymity Table for each Attribute individually}
    \label{tab:3-anon_exploitability_evaluation}
\end{table}

\begin{table}
    \centering
    \begin{tabular}{|c|c|c|c|}
        \hline
         Attributes & Exposure & Inference & Exploitability \\\hline
         \textbf{Age}  & 4-EE & 4-Critical & \textbf{4-Very Easy} \\
         \textbf{Gender} & 4-EE & 3-Severe & \textbf{4-Very Easy}\\
         \textbf{Country}  & 4-EE & 4-Critical & \textbf{4-Very Easy}\\
         Admission Date  & 2-EI & 4-Critical & 3-Easy\\
         Blood Type & 2-EI & 4-Critical & 3-Easy\\\hline
    \end{tabular}
    \caption{Exploitability Assessment for the HIPAA Table for each Attribute individually}
    \label{tab:hipaa_exploitability_evaluation}
\end{table}

\begin{figure}
    \centering
    \includegraphics[width=0.75\linewidth]{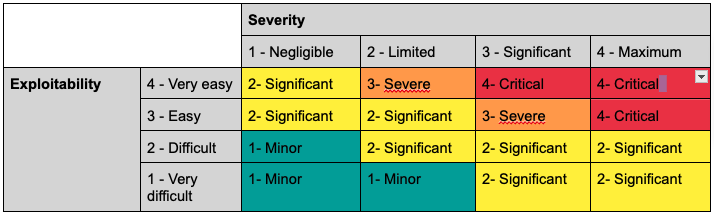}
    \caption{Our Re-identification Risk Scale based on Exploitability and Severity}
    \label{fig:reidentification_scale}
\end{figure}

\begin{table}
    \centering
    \begin{tabular}{|c|c|c|c|}
        \hline
         Description & Exploitability & Severity & Risk Level \\\hline
         \textbf{Re-identification risk based on EE}  & 4-Very Easy & 4-Maximum & \textbf{4-Critical} \\
         \textbf{Re-identification risk based on EI} & 3-Easy & 4-Maximum & \textbf{4-Critical}\\\hline
    \end{tabular}
    \caption{Re-identification Risk Assessment within HIPAA Table}
    \label{tab:hipaa_re-identification_evaluation}
\end{table}

\begin{table}
    \centering
    \begin{tabular}{|c|c|c|c|}
        \hline
         Description & Exploitability & Severity & Risk Level \\\hline
         \textbf{Re-identification risk based on EE}  & 4-Very Easy & 4-Maximum & \textbf{4-Critical} \\
         \textbf{Re-identification risk based on EI} & 3-Easy & 4-Maximum & \textbf{4-Critical}\\\hline
    \end{tabular}
    \caption{Re-identification Risk Assessment within k-anonymity Table}
    \label{tab:3-anon_re-identification_evaluation}
\end{table}

\subsection{Computing the Likelihood}
\label{sec:computing_likelihood}

Based on our proposed exploitability scale, presented in Figure \ref{fig:exploitability_scale}, we are able to calculate the Likelihood, i.e. the exploitability for our two anonymized dataset examples, as shown in Tables \ref{tab:3-anon_attributes_exploitability_evaluation} and \ref{tab:3-anon_exploitability_evaluation} for the k-anonymity dataset and Tables \ref{tab:hipaa_exploitability_evaluation} and \ref{tab:hipaa_attributes_exploitability_evaluation} for the HIPAA dataset. Note that the inference level (4-Critical) is obtained by calculating the DR of the combined quasi-identifiers or the individual attributes. 
For a more conservative assessment, we can assume that an attacker would be able to access all the attributes belonging to the same exposure level at the same time. As indicated in bold in the Tables, the highest risk corresponds to an attacker using the combined attributes "Age", "Gender" and "Country", or the individual attributes "Age" or "Country" to infer "very easily" the "Disease" attribute of an individual, as the exposure level of these attributes are classified as "External Extended". The only difference between Tables \ref{tab:3-anon_exploitability_evaluation} and  \ref{tab:hipaa_exploitability_evaluation} is related to the inference level of the "Blood Type" which is evaluated as severe vs critical thus leading to a difficult vs easy level of exploitability.


\section{Computing the re-identification Risk}
\label{sec:computing_rr}

Based on our proposed re-identification risk scale, presented in Figure \ref{fig:reidentification_scale}, we are able to assess the risk of re-identification for each anonymized dataset, Table \ref{tab:3-anon_re-identification_evaluation} for the k-anonymity dataset and Table \ref{tab:hipaa_re-identification_evaluation} for the HIPAA dataset. 
Both resulting tables show that the risk of re-identification is critical for both anonymized datasets. This is due both to the size of "k" in the k-anonymity Table \ref{tab:3-anon}, which is very small, and to the fact that there is a perfect inference for Group 1. A larger "k" would have provided a better level of security. 



\section{Conclusions}
\label{sec:conclusions}
With the size of the data masking market estimated at \$1.08 billion in 2025 and \$2.14 billion by 2030 \footnote{\url{https://www.mordorintelligence.com/industry-reports/data-masking-market}}, it is important to provide solutions to the enormous challenge posed by the field of database anonymization and the evaluation of such anonymization processing. The goal of this paper is to assist the industry with a practical and ready-to-use methodology, which is fully presented in this paper. 
The methodology introduces new concepts such as severity, exposure and operability, each of which plays a key role in the construction of the methodology. The results are not presented in the form of a cut-and-dried yes or no answer,  but provide more information for assessing risk and interpreting results, and therefore greater precision, with criteria that can be explained and made transparent.


This contribution also has enormous potential in that it can be used to improve anonymization techniques by identifying which attributes or attribute values to focus on in the anonymization process for greater efficiency and to avoid degrading attributes that have little impact on the re-identification risk.

This contribution not only provides a methodology for assessing privacy risks, but also paves the way for designing more accurate anonymization techniques that will help achieve a better trade-off between privacy and data utility.

\bibliographystyle{alpha}
\bibliography{sample}

\newcommand{\etalchar}[1]{$^{#1}$}
\begin{thebibliography}{DEENR12}

\bibitem[BM10]{benitez2010evaluating}
Kathleen Benitez and Bradley Malin.
\newblock Evaluating re-identification risks with respect to the hipaa privacy rule.
\newblock {\em Journal of the American Medical Informatics Association}, 17(2):169--177, 2010.

\bibitem[CDE{\etalchar{+}}23]{carey2023measuring}
CJ~Carey, Travis Dick, Alessandro Epasto, Adel Javanmard, Josh Karlin, Shankar Kumar, Andres Mu{\~n}oz~Medina, Vahab Mirrokni, Gabriel~Henrique Nunes, Sergei Vassilvitskii, et~al.
\newblock Measuring re-identification risk.
\newblock {\em Proceedings of the ACM on Management of Data}, 1(2):1--26, 2023.

\bibitem[DEE10]{dankar2010method}
Fida~Kamal Dankar and Khaled El~Emam.
\newblock A method for evaluating marketer re-identification risk.
\newblock In {\em Proceedings of the 2010 EDBT/ICDT Workshops}, pages 1--10, 2010.

\bibitem[DEENR12]{dankar2012estimating}
Fida~Kamal Dankar, Khaled El~Emam, Angelica Neisa, and Tyson Roffey.
\newblock Estimating the re-identification risk of clinical data sets.
\newblock {\em BMC medical informatics and decision making}, 12:1--15, 2012.

\bibitem[EED08]{el2008protecting}
Khaled El~Emam and Fida~Kamal Dankar.
\newblock Protecting privacy using k-anonymity.
\newblock {\em Journal of the American Medical Informatics Association}, 15(5):627--637, 2008.

\bibitem[EEJAM11]{el2011systematic}
Khaled El~Emam, Elizabeth Jonker, Luk Arbuckle, and Bradley Malin.
\newblock A systematic review of re-identification attacks on health data.
\newblock {\em PloS one}, 6(12):e28071, 2011.

\bibitem[eL18]{CNIL-PIA}
Commission Nationale~Informatique et~Libertés.
\newblock Analyse d'impact relative à la protection des données - privacy impact assessment: Les bases de connaissances, 2018.

\bibitem[EMA18]{EMA}
European Medecines~Agency EMA.
\newblock External guidance on the implementation of the european medicines agency policy on the publication of clinical data for medicinal products for human use.
\newblock Technical Report EMA/90915/2016, EMA, 2018.

\bibitem[EU16]{GDPR2018}
EU.
\newblock {Regulation (EU) 2016/679 of the European Parlament and of the Council of 27 April 2016 on the protection of natural persons with regard to the processing of personal data and on the free movement of such data, and repealing Directive 95/46/EC (General Data Protection Regulation)}, 2016.

\bibitem[GL23]{georgiou2023dpia}
Dimitra Georgiou and Costas Lambrinoudakis.
\newblock Dpia for cloud-based health organizations in the context of gdpr.
\newblock In {\em ECCWS 2023 22nd European Conference on Cyber Warfare and Security}, number~1. Academic Conferences and publishing limited, 2023.

\bibitem[oFA19]{EBIOS}
National Cybersecurity~Agency of~France~(ANSSI).
\newblock Ebios risk manager.
\newblock Technical Report ANssI-pA-048-EN, ANSSI, 2019.

\bibitem[oHH12]{HIPAA}
US~Department of~Health and Human~Services (HHS).
\newblock Guidance regarding methods for de-identification of protected health information in accordance with the health insurance portability and accountability act (hipaa) privacy rule, 2012.

\bibitem[Par14]{WP29Opinion}
Article 29 Data Protection~Working Party.
\newblock Opinion 05/2014 on anonymisation techniques, 2014.

\bibitem[SG20]{seyyar2020privacy}
M~Bas Seyyar and Zeno~JMH Geradts.
\newblock Privacy impact assessment in large-scale digital forensic investigations.
\newblock {\em Forensic Science International: Digital Investigation}, 33:200906, 2020.

\bibitem[SLF17a]{Sondeck2017}
Louis-Philippe Sondeck, Maryline Laurent, and Vincent Frey.
\newblock Discrimination rate: An attribute-centric metric to measure privacy.
\newblock In {\em Annals of Telecommunications}, 2017.

\bibitem[SLF17b]{sondeck2017semantic}
Louis-Philippe Sondeck, Maryline Laurent, and Vincent Frey.
\newblock The semantic discrimination rate metric for privacy measurements which questions the benefit of t-closeness over l-diversity.
\newblock In {\em SECRYPT 2017: 14th International Conference on Security and Cryptography}, volume~6, pages 285--294. Scitepress, 2017.

\end{thebibliography}
\end{document}